\documentclass[aps,prl,twocolumn,english,showpacs,10pt,superscriptaddress]{revtex4-2} \pdfoutput=1
\usepackage{amsmath,bm}
\usepackage{mathtools}
\usepackage{commath}
\usepackage{amssymb}
\usepackage{graphicx}
\usepackage{babel}
\usepackage{cases}
\usepackage[colorlinks=true, citecolor=blue, anchorcolor=green]{hyperref}
\hypersetup{linkcolor=red}
\usepackage{natbib}
\usepackage{floatrow}
\usepackage{dblfloatfix}
\usepackage{xcolor}
\usepackage{braket}

\makeatletter

\newcommand{\Rmnum}[1]{\expandafter\@slowromancap\romannumeral #1@}

\newcommand{\jc}{\text{JC}}

\newcommand{\be}{\begin{equation}}
\newcommand{\ee}{\end{equation}}
\newcommand{\bg}{\begin{aligned}}
\newcommand{\eg}{\end{aligned}}

\makeatother

\begin{document}
\title{ Zeno Regime of Collective Emission: Non-Markovianity beyond Retardation}
\author{Yu-Xiang Zhang}
\email{iyxz@iphy.ac.cn}
\affiliation{Institute of Physics, Chinese Academy of Sciences, Beijing 100190, China}
\affiliation{School of Physical Sciences, University of Chinese Academy of Sciences, Beijing 100049, China}
\affiliation{Hefei National Laboratory, Hefei 230088, China}
\date{\today}

\begin{abstract}
To build up a collective emission, the atoms in an ensemble must coordinate their behavior by exchanging virtual photons. 
We study this non-Markovian process in a subwavelength atom chain coupled
to a one-dimensional (1D) waveguide and find that retardation is not the only cause of non-Markovianity.
The other factor is the memory of the photonic environment, for which a single excited atom needs a finite time,
the Zeno regime,  to
transition from quadratic decay to exponential decay. In the waveguide setup,
this crossover has a time scale longer than the retardation, thus
impacting the development of collective behavior. By comparing a full quantum treatment with an 
approach incorporating only the retardation effect, we find that the field memory effect,
characterized by the population of atomic excitation,
is much more pronounced in collective emissions 
than that in the decay of a single atom. Our results maybe useful for the dissipation engineering of 
quantum information processings based
on compact atom arrays. 
\end{abstract}

\maketitle

It is well known since the 1960s~\cite{Winter:1961aa,Newton:1961aa} that 
the short-time dynamics of an excited atom differs significantly from
the exponential decay based on the Weisskopf-Wigner formalism~\cite{Weisskopf:1930aa}.
The finite memory of the photonic reservoir leads to a growth in the decay rate from zero that is 
quadratic in time~\cite{Wilkinson:1997aa,Crespi:2019aa}.
It has inspired attempts to prevent decay by quickly repeating 
measurements~\cite{Chiu:1977aa,Misra:1977aa,Itano:1990aa,Presilla:1996aa}, i.e., the Zeno effect. 
Actually, the duration of the non-exponential decay, characterized by
the Zeno time  $\tau_Z$, is typically many orders of magnitude shorter
than the lifetime, rendering the field memory effect undetectable. For example, 
an optimal estimation of the 2P-1S transition of the hydrogen atom
reads $\tau_Z\sim 10^{-13}\mathrm{s}$~\cite{Zheng:2008aa}. 
Instead of the decay of a single atom, in this Letter, we study atom ensembles~\cite{Dicke1954},
especially the subwavelength atom arrays, 
where the separation between two adjacent atoms, $d$, is shorter than the resonant wavelength $\lambda$,
and reveal the prominent
memory effect in the Zeno regime.
 
The long-time collective emissions from an atomic ensemble are well described by Lehmberg's formalism based on the Markov
approximation~\cite{Lehmberg:1970aa,Lehmberg:1970ab}. For example, 
it predicts a power-law scaling  $\gamma\propto  N^{-\alpha}$ with $N$ the number of
atoms for the subradiant states of a subwavelength atom array~\cite{Asenjo-Garcia2017,Zhang:2019aa,Zhang:2020ab,Kornovan:2019aa}. 
However, Markovian theories are not able to answer how the
collective behaviors are built up. This process must be non-Markovian
because the atoms are organized by the retarded photon-mediated interactions.
Instead, we may upgrade the Markovian description
minimally by including delayed feedback:  Every atom starts from exponential decay independently
but adjusts its decay rate in response to the signal from
another atom. This physical picture has been analytically studied
for two atoms with the photonic reservoir being 3D free space~\cite{Milonni:1974aa} 
or 1D waveguide~\cite{Gonzalez-Ballestero:2013aa,Sinha:2020aa,Sinha:2020ab}. Retardation effects of waveguide quantum electrodynamics (QED)
are also studied in Refs.~\cite{Dinc:2019aa,Shen:2019aa,Carmele:2020aa,
Arranz-Regidor:2021aa,Trivedi:2021aa,Barkemeyer:2022aa}. 

\begin{figure}[b]
\centering
\includegraphics[width=0.93\textwidth]{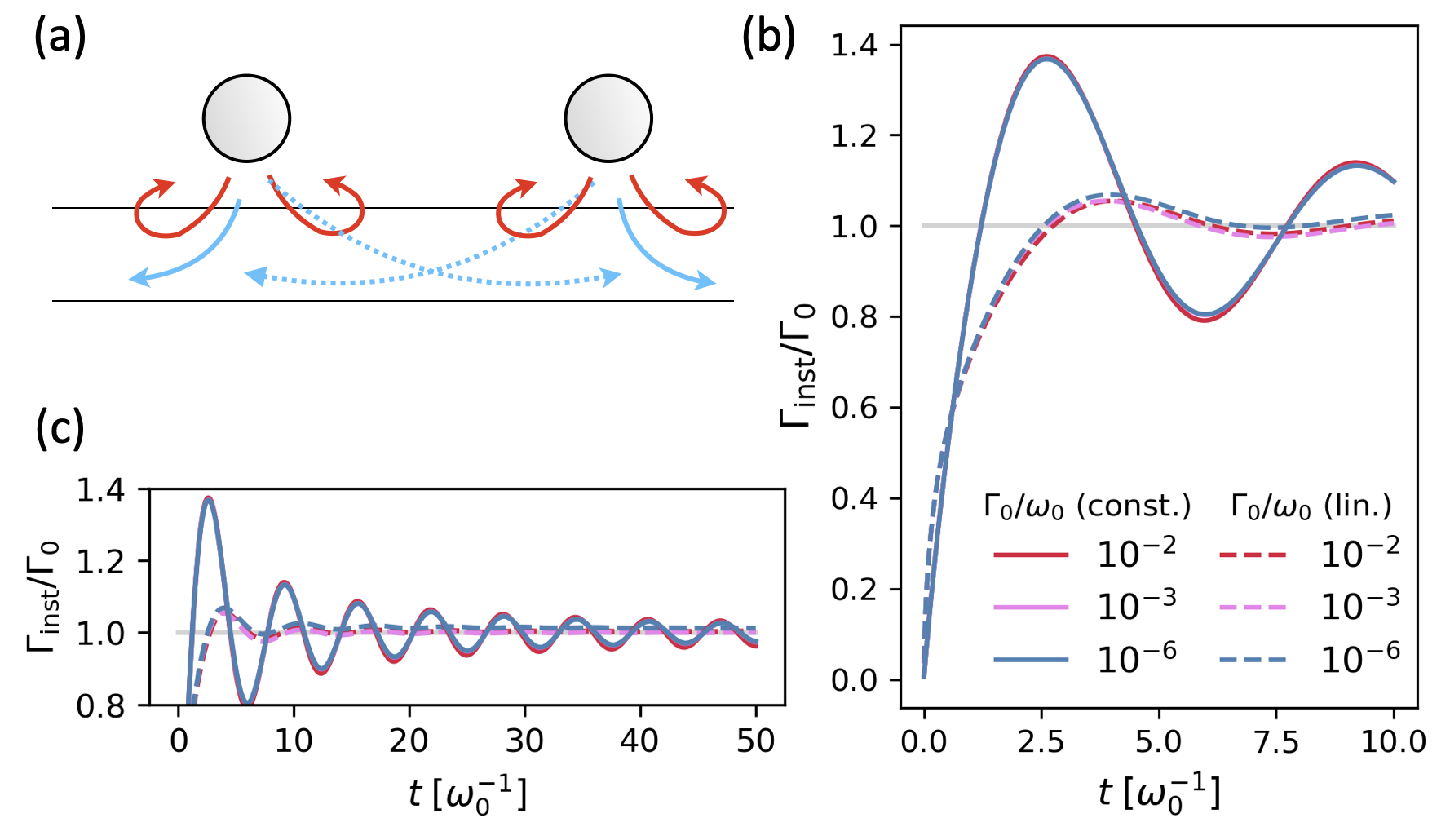}
\caption{(a) Sketch of the idea: For closely separated atoms, development of
collective behaviors by exchanging virtual photons (dotted blue arrows)
and the transition from quadratic to exponential decay of a single atom (Zeno regime caused by the memory effect of the field,
represented by the red back-flow arrows) are highly intertwined. 
(b) The instantaneous emission rate $\Gamma_{\text{inst}}(t)$ of the decay of a single atom 
(in units of $\Gamma_0$). Legend: const-wQED (solid curves), lin-wQED (dashed curves). Values of
$\Gamma_0/\omega_0$ are distinguished by coloring: $10^{-2}$ (red), $10^{-3}$ (pink) and $10^{-6}$ (blue).
(c) Zoom-out view of (b) for $t\leq 50/\omega_0$. }
\label{fig_sys}
\end{figure}

But how does the Zeno regime come into effect? The Zeno time,
exemplified by the
the 2P-1S transition of the hydrogen atom satisfies
$\tau_Z\gg2\pi/\omega_0\sim10^{-15}\mathrm{s}$~\cite{Zheng:2008aa}. In a subwavelength atom array,
the minimal retardation time $t_{\text{retard}}=d/c$, with $c$ the speed of light, 
fulfills that $t_{\text{retard}}<\lambda/c =2\pi/\omega_0 \ll \tau_Z$.
It means that the virtual photons sent from
an atom have already passed many other atoms, building up 
cooperativeness to a certain extent, while an isolated atom has not yet entered exponential decay. 
Thus, the development of  the full collective emission and the reduction to Markovian behavior
are two simultaneous processes highly intertwined. As illustrated in Fig.~\ref{fig_sys}(a),
the above-mentioned retardation-only picture studied in Refs.~\cite{Gonzalez-Ballestero:2013aa,Sinha:2020aa,Sinha:2020ab,Dinc:2019aa,Shen:2019aa,Carmele:2020aa,Arranz-Regidor:2021aa,Trivedi:2021aa,Barkemeyer:2022aa}
does not apply to compact atom ensembles, i.e.,
there are memory effects beyond retardation. 

To show the memory effect beyond retardation, we shall compare the retardation-only picture
with a full quantum treatment. Their difference is displayed by the evolution of
instantaneous decay rates and excited state populations. Our findings indicate that this memory effect in collective emissions is
much more pronounced than that in the decay
of a single atom, making subwavelength atom array a better platform to detect the
non-Markovianity in the Zeno regime.  In this Letter, we shall concentrate on the setup of
waveguide QED. Experimental feasibilities and 
memory effect of free space radiation field will also be discussed.

\paragraph*{Zeno Time of the Waveguide QED.} The waveguide QED setup consists of $N$
two-level atoms with the ground state $\ket{g}$ and the excited state $\ket{e}$, and 
a 1D continuum of bosonic modes. The 
annihilation and generation operator of the waveguide mode with wavenumber $k$ are denoted by
$a_k$ and $a_k^\dagger$, respectively. They satisfy the bosonic commutation relation 
$[a_k, a_{k'}^\dagger]=2\pi\delta(k-k')$.
Hamiltonian of the system is conventionally written in analogy with the multipolar gauge Hamiltonian of quantum optics,
colloquially the  ``$\bm{d}\cdot\bm{E}$'' Hamiltonian~\cite{Roy:2017aa,Chang:2018aa,Sheremet:2023aa}, though
the bosons field may not be photonic, e.g. it could be surface acoustic waves~\cite{Gustafsson:2014aa,Guo:2017aa}
and matter waves~\cite{Vega:2008aa,Krinner:2018aa}, etc.,
\be\label{H1}\bg
H_{M}=&\sum_{i=1}^{N}\omega_0\sigma_i^\dagger\sigma_i
+\int_{-\Lambda}^\Lambda\frac{dk}{2\pi}\omega_k a_k^\dagger a_k \\
&-i\sum_{i=1}^{N}  \int_{-\Lambda}^\Lambda\frac{dk}{2\pi}g_k
\sigma_{i,X}(a_k e^{ikx_i}
-a_k^\dagger e^{-ikx_i})
\eg\ee
where $\sigma_{i}=\ket{g}_i\bra{e}$, $\sigma_{i,X}=\sigma_i+\sigma_i^\dagger$,
$x_i$ denotes the coordinate of atom $i$, $\Lambda$ is the cutoff of wavenumber.
The coupling strength $g_k$ will be specified below. 
Here we assume a linear dispersion relation for the waveguide,
$\omega_k=v_g\abs{k}$, where $v_g$ is the group velocity of the guided modes. 
(The Zeno time of a setup with non-linear dispersion relation is found
qualitatively the same~\cite{sp}.)
The atomic transition frequency $\omega_0$ defines a resonant wavenumber
$k_0=\omega_0/v_g$. We assume $k_0\ll \Lambda$ so that the non-Markovianity induced by
reservoir band edges~\cite{Vats:1998aa} is irrelevant.

The counter-rotating terms of Hamiltonian~\eqref{H1} cannot be ignored 
for short-time dynamics~\cite{Zheng:2008aa}.
Moreover, they are found to lead to non-linearity at the single-photon level~\cite{Jacobs:2023aa}.
Fortunately, theoretical difficulties brought by them can be avoided 
by turning to the gauge introduced by Drummond~\cite{Drummond:1987aa},
which is also called the Jaynes-Cummings (JC) gauge~\cite{Stokes:2019aa,Stokes:2022aa}. 
The transformation from the multipolar gauge to the JC gauge reads
$H_{\jc}=e^{-iS}H_{M}e^{iS}$, where 
\be
S=\sum_{i=1}^{N}\int_{-\Lambda}^{\Lambda}\frac{dk}{2\pi}\frac{g_{k}}{\omega_0+\omega_k}
\sigma_{i,X}(a_{k} e^{ikx_i}+a_{k}^{\dagger} e^{-ik x_i}).
\ee
At the first order of $g_k$, we obtain
\be\label{Hjc}\bg
H_{\jc}\approx &\sum_{i=1}^{N}\omega_0\sigma_i^\dagger\sigma_i
+\int_{-\Lambda}^\Lambda\frac{dk}{2\pi}\omega_k a_k^\dagger a_k \\
&-i\sum_{i=1}^{N}  \int_{-\Lambda}^\Lambda\frac{dk}{2\pi}g_k^{\jc}
(\sigma_{i}^\dagger a_k e^{ikx_i}-\sigma_i a_k^\dagger e^{-ikx_i})
\eg\ee
where $g_k^{\jc}=2g_k\omega_0/(\omega_k+\omega_0)$. 
The neglected terms at the order of
$O(g_k^2)$ give corrections to the atom and waveguide self-energies,
while corrections to their couplings occur at the order of
$O(g_k^3)$~\cite{Zheng:2008aa,Wang:2010aa,Li:2012aa}. 

The absence of counter-rotating terms grants $H_{\jc}$~\eqref{Hjc} a nice property that 
its ground state is identical to that of its free part $\ket{G}=\ket{g_1,g_2\cdots g_N,\emptyset}$, 
an atom-field product state,
where $\emptyset$ denotes the field vacuum. Then, the physical state of exciting an atom
from the overall ground state, $\sigma_i^\dagger\ket{G}$, is also a product state.
This kind of physical states are exactly the initial states interested by us. 
With respect to the multipolar gauge $H_M$, the same physical state is written as
$e^{iS}\sigma_i^\dagger\ket{G}$, which is, however, an entangled state between the atoms and the field.
Recall that initial states in the product form can greatly simply the theoretical analysis
and are essential for theories of open quantum system~\cite{Rivas:2014aa,Breuer:2016aa,Vega:2017aa}.
Thus, we choose to work with $H_{\jc}$ instead of $H_{M}$.

Next, let us specify the coupling strength $g_k$. In the literature of waveguide QED,
a localized atom-field interaction is often assumed~\cite{Shen:2005aa} so that $g_k$ is 
a $k$-independent constant~\cite{Shen:2005aa,Roy:2017aa,Chang:2018aa,Sheremet:2023aa}. It can be spelled by
the Markovian decay rate $\Gamma_0$ as
$g^2_k=\Gamma_0 v_g/2$. Another option for $g_k$ is $g_k\propto \abs{k}^{1/2}$, the same as the multipolar Hamiltonian 
of atoms in free space~\cite{Sheremet:2023aa} (recall that
an atom does not couple to the full displacement field but only its transverse component,
which is a nonlocal field~\cite{Cohen-Tannoudji:1997aa}). In this case, we have
$g^2_k=\Gamma_0v_g\abs{k}/(2k_0)$.
The above two choices for $g_k$ correspond to 
a constant and a linear spectral density, hence will be denoted by
``const-wQED'' and ``lin-wQED'', respectively.

We are now in a position to calculate the Zeno time. Given an initial state $\ket{\Psi_0}$ and 
a Hamiltonian $H$, the Zeno time $\tau_Z$ is defined from
the short-time expansion of the non-decay probability
\be\label{def-zeno}
\abs{\braket{\Psi_0|e^{-iHt}|\Psi_0}}^2
=1-t^2/\tau_Z^2+\cdots.
\ee
The Zeno time is a characterization to the
duration of non-exponential decay, but not an exact measure. Nevertheless, we
substitute $\ket{\Psi_0}=\ket{e,\emptyset}$  and  $H_{\jc}$ with $N=1$
into Eq.~\eqref{def-zeno} and obtain 
\be\label{tau_z}
\tau_Z^{-2}=\begin{cases}
 2\Gamma_0\omega_0/\pi  & \text{const.} \\
 2\Gamma_0 \omega_0\ln(\Lambda/k_0)/\pi & \text{lin.}
\end{cases}
\ee
Remarkably, $\tau_Z$ of const-wQED is independent of the cutoff  $\Lambda$, which is introduced in Eq.~\eqref{H1}.
This is not seen elsewhere. 
The above result implies that $\tau_Z\gg 1/\omega_0$, hence
$\tau_Z\gg \tau_{\text{retard}}$,  is valid if
$\omega_0/\Gamma_0\gg 1$ (for const-wQED) or $\omega_0/\Gamma_0\gg\ln(\Lambda/k_0)$ 
(for lin-wQED).
Such weak atom-field couplings are satisfied commonly.
Zeno time for $N> 1$ is discussed in Ref.~\cite{sp}.

\paragraph{Equation of Motion.} Suppose that the system is initialized with
only one atomic excitation. Note that $H_{\jc}$ preserves the number of excitations, thanks to the
absence of counter-rotating terms. Thus, the evolution is captured by the singly-excited ansatz
\be
\ket{\Psi(t)}=\sum_{i=1}^{N}\alpha_i(t)\sigma_i^\dagger\ket{G}
+\int_{-\Lambda}^{\Lambda}\frac{dk}{2\pi}\beta_k(t)a_k^\dagger\ket{G},
\ee
where $\alpha_i(t)$ and $\beta_k(t)$ are superposition coefficients
to be determined. The Schr{\"o}dinger equation in the interaction picture implies the
integro-differential equation
\be\label{eq-int-dif}
\bg
\frac{d}{dt}\alpha_{i}(t)=
- \sum_{j=1}^{N}\int_{-\Lambda}^{\Lambda} & \frac{dk}{2\pi}  \abs{g_k^{\jc}}^2
\int_0^t d\tau \alpha_j(\tau) \\
&\times e^{ik(x_i-x_j)+i(\omega_k-\omega_0)(t-\tau)}.
\eg\ee
This equation is further transformed into an integral equation and solved numerically~\cite{sp}. 

The above equation will be compared with the following one
embodying only the non-Markovianity caused by retardation~\cite{Sinha:2020aa,Sinha:2020ab}
\be\label{eq-delay}
\frac{d\alpha_{i}(t)}{dt}=-\frac{\Gamma_0}{2}\bigg[\alpha_i(t)
+\sum_{j\neq i}e^{ik_0 r_{ij}}\alpha_j(t-\frac{r_{ij}}{v_g})
\Theta(t-\frac{r_{ij}}{v_g})\bigg]
\ee
where $r_{ij}=\abs{x_i-x_j}$ and $\Theta(t)=1$ for $t>0$ and vanishes otherwise. 
It can be derived from Eq.~\eqref{eq-int-dif} 
via an approximation introduced in Ref.~\cite{Milonni:1974aa}, see also Ref.~\cite{sp}.
Note that while the right-hand-side of Eq.~\eqref{eq-int-dif} incorporates the entire history $\tau\in[0,t]$,
the right-hand-side of Eq.~\eqref{eq-delay} includes only a distance-dependent delay. Ignoring this
delay immediately aligns it with the Markovian effective non-Hermitian Hamiltonian of waveguide QED~\cite{Sheremet:2023aa}.
Hereafter, data produced by Eq.~\eqref{eq-delay} will be labeled by ``retard.''

We shall characterize the non-exponential decay by instantaneous decay rate $\Gamma_{\text{inst}}$ and
population in excited state $P_e(t)$ of the whole chain:
\be
\Gamma_{\text{inst}}(t)\equiv -\frac{d}{dt}\ln P_{e}(t),\quad P_e(t)=\sum_{i=1}^{N}\abs{\alpha_i(t)}^2.
\ee
These two quantities can also be defined for every individual atom in an apparent way. 

\paragraph{Individual decay.}
Let us start from the decay of a single atom. We plot $\Gamma_{\text{inst}}(t)$ in units of $\Gamma_0$
in Figs.~\ref{fig_sys}(b,c) for both const-wQED (solid curves) and lin-wQED (dashed curves).
Either one is calculated with three values of $\Gamma_0/\omega_0$,  $10^{-2}$ (red),
$10^{-3}$ (pink) and $10^{-6}$ (blue). The cutoff is set at $\Lambda/k_0=10^4$. For either const-wQED 
or lin-wQED, curves belonging to the three $\Gamma_0/\omega_0$ almost overlap;
for each value of $\Gamma_0/\omega_0$, $\Gamma_{\text{inst}}(t)$ of lin-wQED 
increases faster at first ($t\lesssim 1/\omega_0$)
but  soon becomes more gradual than that of const-wQED.
The latter increases to roughly $1.4\Gamma_0$ and turns to oscillating around
$\Gamma_0$ with a waning amplitude. 
The non-Markovianity of const-wQED is more pronounced, 
as what we learn from its Zeno time~\eqref{tau_z}.
It is shown in Fig.~\ref{fig_sys}(c) that the oscillation of the curves of const-wQED is still visible
at $t=50/\omega_0$, equivalent to a distance of eight wavelengths for photon propagation. 

\begin{figure}[b]
\raggedright
\includegraphics[width=\textwidth]{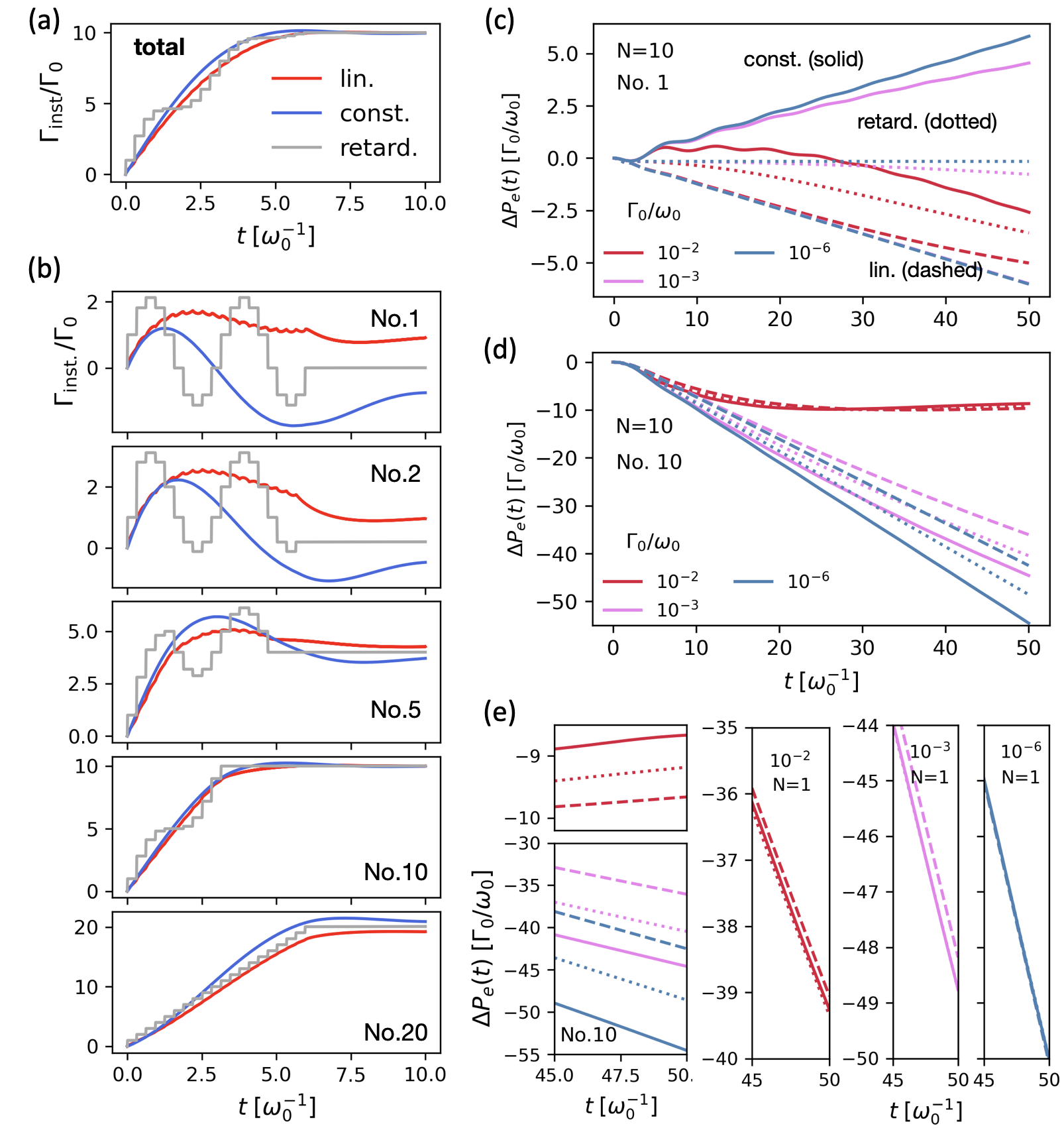}
\caption{The decay of superradiant state  $\ket{\Psi_{k_0}}$. 
(a) The instantaneous decay rate $\Gamma_{\text{inst}}(t)$ for const-wQED (red) and
lin-wQED (blue), both of which are determined by Eq.~\eqref{eq-int-dif}, 
and the retardation-only solution ``retard.'' (grey) given by Eq.~\eqref{eq-delay}. Other parameters: $\Gamma_0/\omega_0=10^{-4}$,
$N=20$, and $d=0.1\pi/k_0$.
(b) The individual instantaneous decay rate of five selected atoms [the legend is the same as in (a)].
The change of individual excited state population $\Delta P_e(t)$ (in units of $\Gamma_0/\omega_0$) 
for (c) atom~1; (d) atom~10.
In (c,d), we have $N=10$, $d=0.1\pi/k_0$ and $\Gamma_0/\omega_0=10^{-2}$ (colored by red)
$10^{-3}$ (pink), and  $10^{-6}$ (blue).
Predictions of const-wQED, lin-wQED and the retardation-only solution ``retard.'' are
plotted by solid, dashed, and dotted curves, respectively.
(e) Left panel: zoom-in view of (d) for $\omega_0t\in [45,\, 50]$;
The three right panels: $\Delta P_e(t)$ of a single atom coupled to the waveguide with
$\Gamma_0/\omega_0=10^{-2}$, $10^{-3}$ and $10^{-6}$, respectively. }
\label{fig_super}
\end{figure}

Although the curves of $\Gamma_{\text{inst}}(t)$ clearly demonstrate the non-exponential decay, 
it is defined as the derivative with respect to time so that
producing the curves requires
a high temporal resolution ($\ll1/\omega_0$) of measuring $P_e(t)$. 
This is of course experimentally challenging. The requirement of temporal resolution might be
relaxed if the non-Markovianity can be manifested by $P_e(t)$ itself, or equivalently,
$\Delta P_e(t)=P_e(t)-P_e(0)$. Unfortunately, we will see
in Fig.~\ref{fig_super}(e) that it is not the case for $N=1$: $\Delta P_e(t)$ is 
averaged out to the memory-less Markovian result quickly.
But fortunately, it would be possible for subwavelength atom arrays ($N>1$). 

\paragraph*{Superradiance.} We consider a chain of atoms 
initialized in the timed-Dicke state 
\be\label{state}
\ket{\Psi_k}=\frac{1}{\sqrt{N}}\sum_{j=1}^{N}e^{ikx_j}\sigma_j^\dagger\ket{G}.
\ee
This kind of states are experimentally accessible~\cite{Rohlsberger:2010aa,Corzo:2019aa}.
State~\eqref{state} with $k=\pm k_0$ are the single-photon superradiant state~\cite{Scully:2006aa}. 
We substitute $\ket{\Psi_{k_0}}$ with $N=20$, $\Gamma_0/\omega_0=10^{-4}$, and atom-atom separation $d=0.1\pi/k_0$
(see results of $d=0.5\pi/k_0$ in~\cite{sp}) into
into Eqs.~\eqref{eq-int-dif} and~\eqref{eq-delay} and show the results of $\Gamma_{\text{inst}}(t)$
in Fig.~\ref{fig_super}(a) for $t\leq 10/\omega_0$. Curves of Eq.~\eqref{eq-int-dif} (red for lin-wQED and blue for
const-wQED)
show continuous growth
while that of Eq.~\eqref{eq-delay} (grey) gives a step-like increase. They agree well after $t\approx 6/\omega_0$. 

Next, we pick five atoms, No.~1, 2, 5, 10,  20 ($x_i<x_j$ if $i<j$), and plot the
individual instantaneous decay rate, $\Gamma_{j,\text{inst}}(t)=-d(\ln \abs{\alpha_j}^2)/dt$
in Fig.~\ref{fig_super}(b). It shows that the atoms decay at different rates.
Atom~1 decays slowly while Atom~20 accelerates to $20\Gamma_0$ (the opposite is obtained if we
choose $\ket{\Psi_k}$ with $k=-k_0$).
The three curves (const-wQED, lin-wQED and retard.)  agree better
for atoms in the middle, i.e., atom~10. 
In particular, for atom~1, significant derivations between three curves of $\Gamma_{1,\text{const}}$
are visible: the grey curve (retard.) shows two cycles of emission and absorption, 
the red curve (const-wQED) shows only emission while
absorption is dominant for the blue curve (lin-wQED). 

Such discrepancy inspires us to look at the change of individual population
$\Delta P_{e}=\abs{\alpha_j(t)}^2-\abs{\alpha_j(0)}^2$, where $\abs{\alpha_j(0)}^2=1/N$ for state~\eqref{state}. 
In Fig.~\ref{fig_super}(c), we plot it for atom~1 of a shorter chain ($N=10$) within 
a longer time window $t\leq 50/\omega_0$ .
To compare with Fig.~\ref{fig_sys}, we apply the same three 
values of  $\Gamma_0/\omega_0$ and the same coloring as in Fig.~\ref{fig_sys}.
Figure~\ref{fig_super}(c) shows that for $\Gamma_0/\omega_0=10^{-3}$ (pink) and $10^{-6}$ (blue)
there is a gap of $\lesssim 10\Gamma_0/\omega_0$ between const-wQED (solid curves) and lin-wQED (dashed curves), 
while the predictions of
Eq.~\eqref{eq-delay} (dotted curves) are roughly in the middle.
For stronger atom-waveguide coupling $\Gamma_0/\omega_0=10^{-2}$, 
the curves of const-wQED (red solid) and  that of Eq.~\eqref{eq-delay} (red dotted) have a tendency toward getting closer.
We also plot $\Delta P_{e}(t)$ for the last atom (No.~10) in Fig.~\ref{fig_super}(d).
It shows gaps between const-wQED (solid) and lin-wQED (dashed) of roughly the same scale as atom~1, 
except for the case of $\Gamma_0/\omega_0=10^{-2}$.  A zoom-in view for  $\omega_0 t\in [45, 50]$ is shown 
in the left-most panel of Fig.~\ref{fig_super}(e).

To compare, we plot $\Delta P_e(t)$ for the case of $N=1$ in the right three panels of Fig.~\ref{fig_super}(e), 
each for one choice of $\Gamma_0/\omega_0$.
We find that predictions of the three models (solid, dashed and dotted curves) 
are almost indistinguishable for  $\omega_0 t\in [45, 50]$.
Thus, in terms of $\Delta P_e(t)$, the non-Markovian effect in the Zeno regime is much 
more prominent in collective emissions than in the decay of a single atom.
And it is reasonable to conclude that the requirement of temporal resolution is significantly relaxed to the level of
$\lesssim 10/\omega_0$.

\begin{figure}[tb]
\centering
\includegraphics[width=0.88\textwidth]{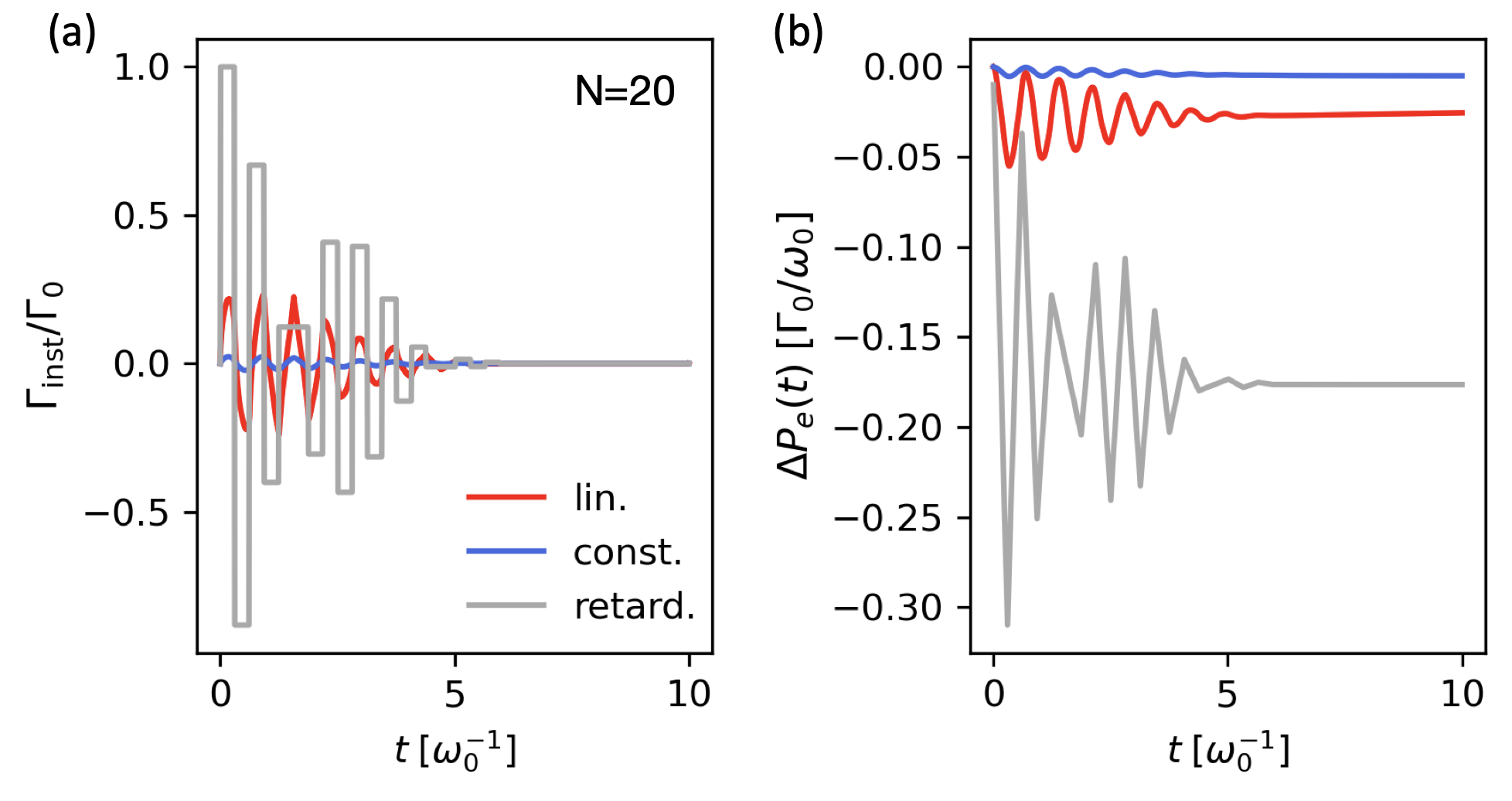}
\caption{The decay of the subradiant state $\ket{\Psi_{\text{sub}}}$ with $N=20$. (a) The instantaneous decay rate $\Gamma_{\text{inst}}(t)$ (in units of $\Gamma_0$). The other
parameters and the legend are the same as in Fig.~\ref{fig_super}(a).
(b) The change of population on the excited state $\Delta P_e(t)$ (in units of $\Gamma_0/\omega_0$).
}
\label{fig_sub}
\end{figure}

\paragraph*{Subradiance.}
The effective Hamiltonian of waveguide QED defines a subradiant eigenstate approximated by 
$\ket{\Psi_{\text{sub}}}=(\ket{\Psi_k}-\ket{\Psi_{-k}})/\sqrt{2}$ with $kd=\pi N/(N+1)$~\cite{Asenjo-Garcia2017,Zhang:2019aa,Zhang:2020ab}. 
We suppose the same parameters as in Fig.~\ref{fig_super}(a,b) and 
plot $\Gamma_{\text{inst}}(t)$ for a chain initialized in $\ket{\Psi_{\text{sub}}}$ in Fig.~\ref{fig_sub}(a). 
In all cases, the subradiance is built through quick oscillations between emissions and absorptions.
But the amplitudes are different: The piecewise curve predicted by Eq.~\eqref{eq-delay} has the largest amplitudes.
In Fig.~\ref{fig_sub}(b), we plot $\Delta P_e(t)$ of the whole chain. It also shows apparent
relative discrepancies between the three predictions. 

\paragraph*{Discussions.}  It is of fundamental interest to extend the theory to atoms in free space.
However, a controversial issue is that on which Hamiltonian all calculations should be based.  
Most works chose the Coulomb gauge ($\bm{A}\cdot\bm{p}$ interaction) disregarding the $\bm{A}^2$ 
term, see, e.g., Refs.~\cite{Moses:1973aa,Wodkiewicz:1976aa,Seke:1988aa,Facchi:1998aa,Berman:2010aa}. 
Taking the counter-rotating terms into account, it has been found that $\tau_Z^{-2}\sim \ln\Lambda$~\cite{Zheng:2008aa},
the same as lin-wQED.
Thus, we expect the same non-Markovianity, or perhaps
even more pronounced, because the resonant dipole-dipole interaction in free space
diverges as $1/r^3$ for short distances, resulting in strong photon blockade~\cite{Cidrim:2020ub}. 
But recently Hamiltonians of quantum optics is revisited by causal perturbation theory~\cite{Marzlin:2018aa}. 
In this sense, non-Markovianity beyond retardation 
might be viewed as a probe to determine which theory better captures the true physics.

For experimental tests, our plots show that temporal resolution at the scale of $\lesssim 10/\omega_0$ is favorable.
Among various platforms of waveguide QED, superconducting circuits have the highest coupling efficiency~\cite{Sheremet:2023aa}
and fabricating the transmon qubits into a subwavelength chain is straightforward~\cite{Mirhosseini:2019aa}. 
The transition frequency $\omega_0$ is in the GHz regime so that temporal resolution at nanosecond is sufficient. 
Subwavelength atom arrays can also be realized by trapping Sr atoms
in optical lattices~\cite{Olmos:2013aa}. The wavelength of $\prescript{3}{}{P}_0- \prescript{3}{}{D}_1$ 
transition is $2.6\mathrm{\mu m}$ so the
temporal resolution should be $\lesssim 10\mathrm{fs}$. Scenarios where the boson fields are
surface acoustic waves~\cite{Gustafsson:2014aa,Guo:2017aa}
and matter waves~\cite{Vega:2008aa,Krinner:2018aa} need further studies.

\paragraph*{Conclusions.}
We have studied the Zeno regime of the decay of
subwavelength atom arrays coupled to a 1D waveguide. 
Non-markovianity beyond retardation, characterized by instantaneous decay rates
and population in the excited states, is addressed by comparing 
the full quantum treatment Eq.~\eqref{eq-int-dif} with Eq.~\eqref{eq-delay},
which includes only the retardation effect.
Specifically, the evolution of excited state population (in the single-photon superradiant state) 
manifests reservoir memory effect with a significantly relaxed temporal resolution.
Our results might be useful for protecting 
the quantum information stored in compact atom ensembles~\cite{Asenjo-Garcia2017,Manzoni:2018aa}
via dissipation engineering~\cite{Harrington:2022aa},
and studying the correlated noise in quantum computing processors~\cite{Singh:2023aa}, etc.
For future works, one may explore such possibilities using the theoretical tools
of non-Markovian open systems~\cite{Rivas:2014aa,Breuer:2016aa,Vega:2017aa}.

\begin{acknowledgements}
The author thanks Klaus M{\o}lmer for careful reading of the manuscript and for providing valuable comments.
The author acknowledges the financial support from CAS Project for Young Scientists in Basic Research (YSBR-100),
National Natural Science Foundation of China (Grant No.~12375024),
Innovation Program for Quantum Science and Technology (Grant No.~2-6), and the startup grant of
IOP-CAS.
\end{acknowledgements}

\bibliography{supsub}


\end{document}


\title{ SUPPLEMENTAL MATERIAL}
\author{Yu-Xiang Zhang}
\email{iyxz@iphy.ac.cn}
\affiliation{Institute of Physics, Chinese Academy of Sciences, Beijing 100190, China}
\affiliation{School of Physical Sciences, University of Chinese Academy of Sciences, Beijing 100049, China}
\affiliation{Hefei National Laboratory, Hefei 230088, China}
\date{\today}

\begin{abstract}
The supplemental material includes: Sec.~A, a discussion of Zeno time for a chain of $N>1$ atoms in a timed-Dicke state;
Sec.~B, Zeno time of a waveguide with non-linear dispersion relation;
Sec.~C, more details of Eq.~(7) of the main text; Sec.~D, a derivation of Eq.~(8) of the main text;
Sec.~E, instantaneous decay rates of a chain of ten atoms with the spacing $d=0.5\pi/k_0$.
\end{abstract}

\maketitle

\subsection{A. Zeno time for $N>1$}\label{Zeno-N}
In this section, we calculate the Zeno time $\tau_{Z,N}$ for N atoms initialized in 
the timed-Dicke state $\ket{\Psi_{k}}$, i.e., state~(10) of the main text. Substituting the state into the
definition~(4) of the main text, we obtain
\be\label{seq-zeno}
\bg
\tau_{Z,N}^{-2} &= \big(\braket{\Psi_k | H | \Psi_k}\big)^2-\braket{\Psi_k | H^2 | \Psi_k } \\
& =\frac{1}{2N}\sum_{i,j=1}^{N} \int_{-\Lambda}^{\Lambda}
\frac{dq}{2\pi} (g_q^{\jc})^2 e^{i(q-k)(x_i-x_j)} \\
& =\tau_Z^{-2}+\frac{1}{2N}\sum_{i\neq j} e^{-ik(x_i-x_j)} \int_{-\Lambda}^{\Lambda}
\frac{dq}{2\pi} (g_q^{\jc})^2 e^{iq(x_i-x_j)}
\eg\ee
where the contribution of the diagonal terms $(i=j)$
is exactly $\tau_Z$, the Zeno time of a single atom evaluated in Eq.~(5) of the main text. 

It is conceivable that the contribution from off-diagonal terms $(i\neq j)$ is less divergent
in terms of the cutoff $\Lambda$, because they have a phase oscillating with $q$
in the integrand of Eq.~\eqref{seq-zeno}. For example, the scale of the integrals of
lin-wQED reads
\be
\sim \int_0^{\Lambda} dq \bigg(\frac{\sqrt{q}}{1+q}\bigg)^2 e^{iq r_{ij}},
\ee
where $r_{ij}=k_0 \abs{x_i-x_j}$. It diverges in the form of $\ln\Lambda$ if $i=j$ but convergences otherwise.
Thus, the off-diagonal terms are negligible. 

However, the above argument does not apply to the case of const-wQED
because the diagonal terms do not divergent with $\Lambda$ at all.
Scales of the integrals of const-wQED are estimated as
\be
\sim \int_0^\Lambda dq \bigg(\frac{1}{1+q}\bigg)^2 e^{iq r_{ij}}.
\ee
It converges with increasing $\Lambda$ even for the case of $i=j$. 
By a few lines of calculations we obtain that
\be\bg
&\lim_{\Lambda\rightarrow \infty}\int_0^{\Lambda} dq \bigg(\frac{1}{1+q}\bigg)^2 e^{iq r_{ij}} \\
=& 1-r_{ij}e^{-ir_{ij}}\big[\mathrm{si}(-r_{ij})+i\mathrm{ci}(r_{ij})\big]. 
\eg\ee
The above integral shows a constant term regardless of $r_{ij}$ and an oscillating term.
It turns that the relation between $\tau_{Z,N}$ and $\tau_Z$ is subtle. If the initial state
has a wavenumber $k=O(1)$, the summation of  $e^{ik(x_i-x_j)}$ in Eq.~\eqref{seq-zeno} 
will wash out the off-diagonal terms. But if $k=O(N^{-1})$ in particular $k=0$, a linear scaling of
$\tau_{Z, N}^{-2}$, hence $\tau_{Z,N}\sim N^{-1/2}$ might be possible. That is,
a long chain may not favor a longer $\tau_Z$.

Nevertheless, $\ket{\Psi_k}$ with $k=0$ is an subradiant state
of waveguide QED~\cite{Zhang:2019aa}. It is not pertinent to the superradiant states 
that are mostly interested by us. And a shorter chain is more accessible in practice.

\subsection{B. Zeno time for waveguide with non-linear dispersion relation}\label{sin}
In Ref.~\cite{Zhang:2021aa} the author and colleagues studied 
a waveguide QED setup realized by superconducting circuits, where an artificial atoms 
of finite size ($\delta x$) couples to a transmission line. 
The transmission line is modeled as a chain of coupled LC loops,
where each loop has inductance $l\delta_z$ and conductance $c\delta_z$. We refer the readers to
the supplemental material of Ref.~\cite{Zhang:2021aa} for details of the circuits.
In short, the take home message is that we have $g_k\propto \sqrt{w_k}$ and
\be
\omega_k= J\abs{\sin \frac{k}{2}}
\ee
where $J=\frac{2}{\sqrt{lc}\delta_z}$ plays the role of energy cutoff, and the wavenumber $k$ is defined in
the Brillouin zone $[-\pi, \pi]$. 

To evaluate the Zeno time, we define the group velocity at the 
resonant frequency $\omega_0$, $v_g=\frac{J}{2}\cos(\frac{k}{2})$. Then, the Zeno time of a single atom
reads
\be
\bg
\tau_Z^{-2} & = \frac{4\Gamma_0 v_g \sin(k_0/2}{\pi}\int_{0}^{\pi/2} dx\,
\frac{\sin(x)}{\big[\sin(\frac{k_0}{2})+\sin(x)\big]} \\
&= \frac{4\Gamma_0 v_g a}{\pi(1-a^2)}\big[ -1 + \frac{1}{\sqrt{1-a^2}} 
\ln \frac{1+\sqrt{1-a^2} + a}{1-\sqrt{1-a^2} + a} \big],
\eg\ee
where $a=\sin(k_0/2)$. Now we assume that the resonant frequency is far from the cutoff, i.e., $k_0 \ll \pi$.
It leads to the approximation that $2v_g a\approx \omega_0$ so that 
\be
\tau_Z^{-2}\approx \frac{2\Gamma_0\omega_0}{\pi}\big[ \ln(2\Lambda+1)-1\big]
\ee
where $\Lambda=a^{-1}=J/\omega_0$. That is, the Zeno time of this model is qualitatively the same as
the Zeno time of lin-wQED.

\subsection{C. Kernel of the Integro-Differential Equation}\label{kernel}
In our numerical calculations, the integro-differential equation~(7) of the main text 
is transformed to an integral equation 
\be
\alpha_i(t)-\alpha_i(0)=i\frac{2\Gamma_0}{\pi}\int_0^t d\tau [A_{ij}(t-\tau)]\alpha_j(\tau),
\ee
where the kernel is specified in the below.
To determine the kernels in the above integral equations,  we
introduce the unit less notations $\phi=\omega_0(t-\tau)$, $r_{ij}=k_0\abs{x_i-x_j}$ and
a short hand replacement  $\Lambda/k_0\rightarrow\Lambda$. Then we have that
in the constant case,

\be\bg
A^{\text{const.}}_{ij}(\phi)  & =\int_0^\Lambda dx \frac{\cos(x r_{ij})}{(1+x)^2}\frac{1-e^{i(1-x)\phi}}{x-1}  \\
&= \frac{1}{4}(I_1 - I_2 -2I_3),
\eg\ee
and in the linear case,
\be\bg
A^{\text{lin.}}_{ij}(\phi)&= \int_0^\Lambda dx \frac{x\cos(x r_{ij})}{(1+x)^2}\frac{1-e^{i(1-x)\phi}}{x-1}  \\
&=\frac{1}{4}(I_1- I_2 + 2I_3);
\eg\ee
where the abbreviated terms are
\begin{widetext}
\begin{subequations}
\be\label{sub-I1}
(I1)\;  \int_0^\Lambda dx \frac{\cos(x r_{ij})}{x-1}(1-e^{i(1-x)\phi}) 
= \bigg[ \cos(r_{ij})\ci(xr_{ij})-\sin(r_{ij})\si(xr_{ij})
-\frac{e^{ir_{ij}}}{2} \mathrm{csi}(x r^{-}_{ij})  -\frac{e^{-ir_{ij}}}{2}\mathrm{csi}^{*}(xr^{+}_{ij})\bigg]
\bigg\vert_{-1}^{\Lambda-1}
\ee
\be\bg
(I2)\; &\int_0^\Lambda dx \frac{\cos(x r_{ij})}{x+1}(1-e^{i(1-x)\phi})\\
= & \bigg[\cos(r_{ij})\ci(xr_{ij})+\sin(r_{ij})\si(xr_{ij}) 
-\frac{e^{i(\phi-r^{-}_{ij})}}{2} \mathrm{csi}(x r^{-}_{ij}) 
-\frac{e^{i(\phi+r^{+}_{ij})}}{2}\mathrm{csi}^{*}(xr^{+}_{ij}) \bigg]
\bigg\vert_{1}^{\Lambda+1}
\eg\ee
\be\bg
(I3)\; & \int_0^\Lambda dx \frac{\cos(x r_{ij})}{(1+x)^2}[1-e^{i(1-x)\phi}] 
=  \frac{\cos(\Lambda r_{ij})}{\Lambda+1}(e^{-i(\Lambda+1)\phi}-1)
+1-e^{i\phi} \\
&\qquad -\bigg\{r_{ij}\big[\cos(r_{ij}) \mathrm{si}(x r_{ij}) - \sin(r_{ij})\mathrm{ci}(xr_{ij})\big] +\frac{i}{2}r_{ij}^{-} 
e^{i(2\phi-r_{ij})}\mathrm{csi}(xr_{ij}^{-})
-\frac{i}{2}r_{ij}^{+} e^{i(2\phi+r_{ij})}\mathrm{csi}^{*}(xr_{ij}^{+}) \bigg\}\bigg\vert_{1}^{\Lambda+1}
\eg\ee
\end{subequations}
\end{widetext}
where $r_{ij}^{\pm}=r_{ij}\pm\phi$, and we have introduced
$$\mathrm{csi}(x)\equiv \ci(x)+i\si(x)\equiv-\int_{x}^{\infty}dt\frac{e^{it}}{t},$$
and its complex conjugate $\mathrm{csi}^*(x)$.

The population of the bosonic field excitation 
$$N_{b}(t)=\int_{-\Lambda}^{\Lambda}\frac{dk}{2\pi}\abs{\beta_k(t)}^2$$
is obtained from the atomic amplitudes by
\be
N_b(t)=\sum_{i,j=1}^{N}
\int_0^t d\tau \int_0^t d\tau' B_{ij}(\tau-\tau')\alpha_i(\tau)\alpha_j(\tau'),
\ee
where the kernel is defined by
\be
B_{ij}(t-\tau)=\int_{-\Lambda}^{\Lambda} \frac{dk}{2\pi}\abs{g_k^{\jc}}^2 e^{-ik(x_i-x_j)-i(\omega_0-\omega_k)(t-\tau)}
\ee
For the constant case, the kernel of photon population is evaluated as
\be \bg
\; B_{ij}^{\text{const.}}(\phi)  & =\frac{\Gamma_0\omega_0 }{\pi}
\int_0^\Lambda dx\frac{1}{(1+x)^2}2\cos(xr_{ij})e^{i(x-1)\phi}  \\
&=  \frac{\Gamma_0\omega_0e^{-i\phi}}{\pi}\bigg [
-\frac{2\cos(xr_{ij})}{1+x}e^{ix\phi}\big\vert_0^\Lambda \\
&\quad +\sum_{s=\pm} i \phi_s e^{-i\phi_s} 
\mathrm{csi}(x\phi_s)\big\vert_1^{\Lambda+1} \bigg]
\eg\ee
where $\phi_{\pm}=\phi\pm r_{ij}$;
for the linear case, we have
\be\bg
B_{ij}^{\text{lin.}}(\phi)  &=\frac{\Gamma_0\omega_0 }{\pi}
\int_0^\Lambda dx\frac{x}{(1+x)^2}2\cos(xr_{ij})e^{i(x-1)\phi}  \\
&=  \sum_{s=\pm} e^{-i \phi_s} \mathrm{csi}(x\phi_s)\big\vert_1^{\Lambda+1} - B_{ij}^{\text{const}}(\phi).
\eg\ee

We solved both the population of atomic excitation and the number of photonic excitations. The sum of them,
which should be one in theory, is used to renormalize the numerical solutions.

\subsection{D. Derivation of Eq.~(8) of the main text}\label{delay}

Equation~(8) of the main text goes beyond the Markovian equation of motion (governed by the non-Hermitian
effective Hamiltonian, in the absence of the quantum jumps) minimally. It shares 
a properties with the Markovian formalism that it is gauge invariant~\cite{Stokes:2022aa}.
We can derive it from any choice of Hamiltonian. (Hamiltonian is of course gauge-dependent.) 

In order to derive Eq.~(8), the trick firstly introduced in Ref. ~\cite{Milonni:1974aa} 
is to substitute the following approximation
\be
\int\frac{dk}{2\pi} f(k) e^{i(k_0-k)x} \approx f(k_0)\delta(x)
\ee
into Eq.~(7) of the main text. The above approximation is valid
if the function $f(k)$ changes slowly with $k$. In the large $\Lambda$ limit, it leads to
\be\label{eq-int-dif}
\bg
\frac{d}{dt}\alpha_{i}(t)=
- \sum_{j=1}^{N}  \frac{1}{v_g}  \abs{g_{k_0}^{\jc}}^2
\int_0^t d\tau \alpha_j(\tau) e^{ik_0\abs{x_i-x_j}} \\
\delta(-\frac{\abs{x_i-x_j}}{v_g}+t-\tau).
\eg\ee
Note that now only the coupling strength at the resonant momentum $k_0$
shows up. Thus there is no difference to base the derivation on either const-wQED or lin-wQED.
The above equation reduces to Eq.~(8) of the main text by using $\Gamma_0=2 g^2_{k_0}/v_g$.

\begin{figure}[t]
\centering
\includegraphics[width=0.99\textwidth]{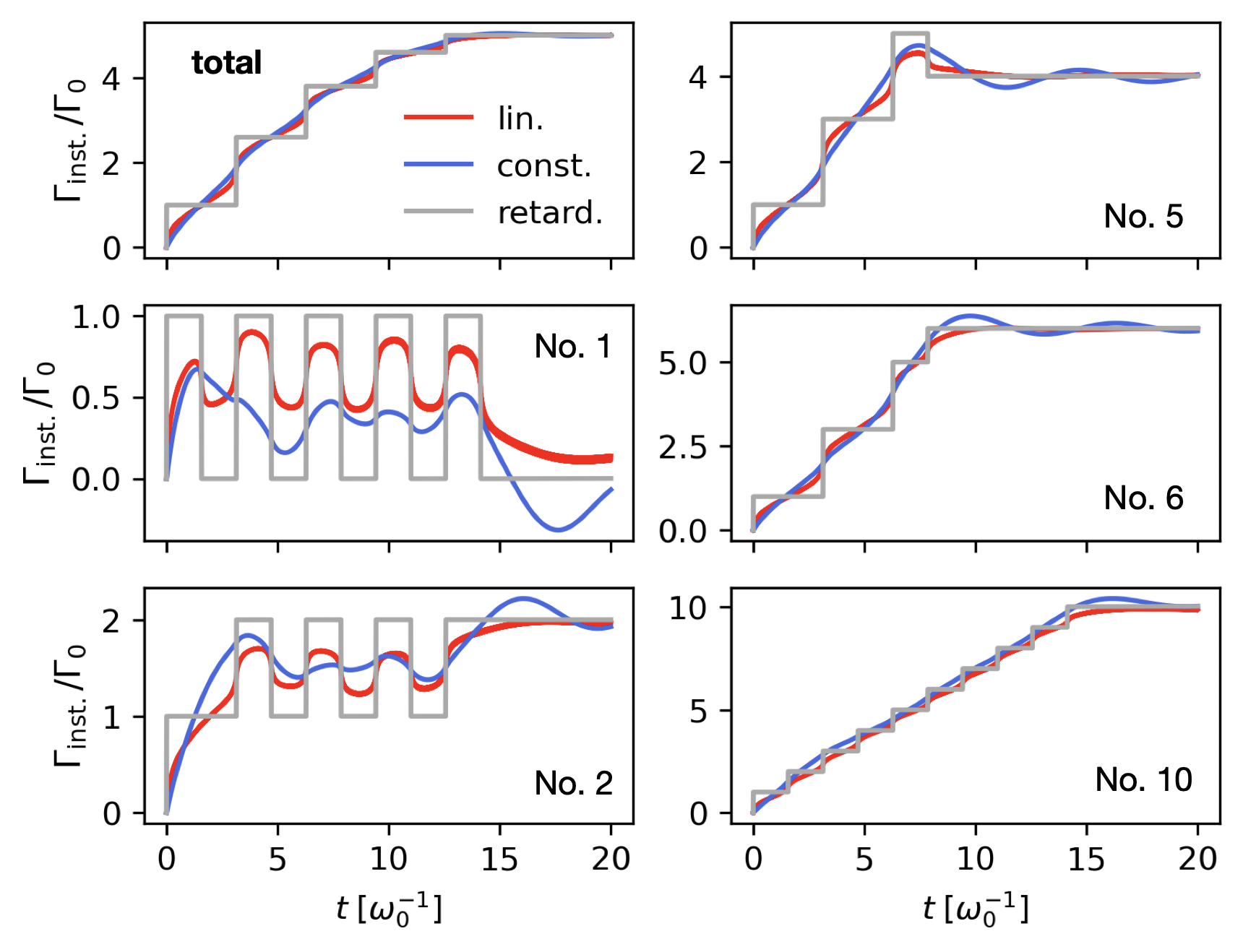}
\caption{
The instantaneous decay rate $\Gamma_{\text{inst}}(t)$ (in the unit of $\Gamma_0$) 
of a chain of 10 atoms initialized in $\ket{\Psi_{k_0}}$ separated by $d=0.5\pi/k_0$.
}
\label{sp_super}
\end{figure}

\subsection{E. Results of $d=0.5\pi/d$}\label{plot}
Results presented in the main text belong to subwavelength arrays with the atom-atom separation $d=0.1\pi/k_0$.
The field memory effect will be diminished if $d$ becomes larger. To see it we consider a chain of 10 atoms
with $d=0.5\pi/d$ and $\Gamma_0/\omega_0=10^{-4}$, and suppose the chain is initialized 
in the single-photon superradiant state $\ket{\Psi_{k_0}}$. We show in Fig.~\ref{sp_super} the instantaneous decay 
rate of the whole chain and of five atoms, No.~1, 2, 5, 6, 10.
It can be seen that the differences between const-wQED, lin-wQED and Eq.~(8) of the main text
are less prominent than the case of $d=0.1\pi/k_0$.

\bibliography{supsub}
